\documentclass[twocolumn,showpacs,preprintnumbers,amsmath,amssymb,aps,prl]{revtex4}
\usepackage{graphicx}
\begin{document}

\title{
Casimir Effect in Active Matter Systems 
} 
\author{
D. Ray$^{1,2}$, C. Reichhardt$^1$, and  C. J. Olson Reichhardt$^{1}$} 
\affiliation{
$^1$Theoretical Division,
Los Alamos National Laboratory, Los Alamos, New Mexico 87545 USA\\ 
$^2$Department of Physics, University of Notre Dame, Notre Dame,
Indiana 46556, USA
} 

\date{\today}
\begin{abstract}
We numerically examine run-and-tumble active matter particles in 
Casimir geometries composed of two finite parallel walls. 
We find that there is an attractive force between the two walls 
of a magnitude that increases with increasing run length.
The attraction
exhibits an unusual exponential dependence on the wall separation, 
and it arises due to a depletion of swimmers in the region between the
walls by a combination of the motion of the particles along
the walls and a geometric shadowing effect.
This attraction is robust as long as the wall separation is comparable to or
smaller than the swimmer run length, and is only slightly reduced by the 
inclusion of steric interactions between swimmers. 
We also examine other geometries and find regimes in which there is a
crossover from attraction to repulsion between the walls as a function of 
wall separation and run length.
\end{abstract}
\maketitle

\vskip2pc
The Casimir geometry consists of two finite parallel plates 
placed at a fixed distance from each other that experience an
attraction due to the confinement of fluctuations in the media
between the plates \cite{1}.
In the original 
Casimir effect calculation, 
electromagnetic vacuum fluctuations 
produce an attractive force between 
two metal places in a vacuum.
Much later, Casimir forces
were experimentally measured \cite{2},
and they have recently been studied 
in 
a variety of systems in order to understand how to 
control 
their magnitude \cite{N} 
or 
polarity \cite{3}.   
Confined classical fluctuations can produce the so-called
critical Casimir effect \cite{4,6,5},
as first proposed by Fisher and de Gennes near critical demixing in bulk 
mixtures. 
The critical
Casimir effect
has been directly measured \cite{7} and studied in various colloidal 
systems where 
it can produce colloidal aggregation \cite{7,8}. 
Casimir type effects have also been studied
in granular media, where attractive forces 
arise between 
objects or plates placed in vibrated or flowing sand \cite{9}.   
Critical Casimir effects have also been proposed to occur 
near percolation
thresholds \cite{10} and 
in biological
systems such as near fluctuating cellular membranes \cite{11}. 
The ability to enhance or control such forces 
can 
lead to a wide variety of applications in self-assembly, 
particle transport, and the creation of novel devices. 

Strong fluctuations appear in active matter or self-driven particle
systems \cite{12} 
such as
swimming bacteria 
undergoing run-and-tumble dynamics \cite{13,14}. 
Recently a number of non-biological active matter systems have been 
realized experimentally, 
including artificial swimmers \cite{15}, 
self-driven 
colloids \cite{16,17,18}, 
or light-activated colloidal particles performing a directed
random walk
\cite{17,18,23}. 
Studies of interacting
particles undergoing run-and-tumble or active Brownian motion 
show a phase separation phenomenon 
at large run length or high density 
in which the particles form dense
regions separated by a dilute active gas 
\cite{20,21,22,23}. 
Monodisperse active particles, 
such as self-propelled disks, can form 
intermittent dense patches with crystalline order,
termed ``living crystals'' \cite{18}.
It has been shown that run-and-tumble dynamics and active Brownian motion 
can be mapped on to each other so that results obtained with one class of
system should be generalizible to the other \cite{24}.

Here we address the question of whether a Casimir-like attractive force
arises between two plates placed in a bath of active particles, and whether
such a force can be controlled by changing the plate geometry.
There is already some evidence that active matter
systems, particularly run-and-tumble particles,
can induce forces on objects. 
One example is the ratchet effect 
observed for swimming bacteria and run-and-tumble particles 
in the presence of an array of asymmetric funnels \cite{14,26,27,28,30}.
Here, when the active particles
run along the walls of the funnel they can escape through
easy direction of the funnel or become trapped in a funnel tip.
While in contact with a wall, their continued swimming produces
forces on the funnel walls.
Such induced forces were more clearly
demonstrated in systems 
where an untethered asymmetric sawtooth gear rotates in a preferred
direction when placed in a bacterial bath \cite{32,33},
but only if the bacteria are actively swimming. 
In simulations, 
asymmetric objects 
placed in an active matter
bath are pushed to produce
an active matter-driven micro-shuttle \cite{31,34}.               

In the Casimir geometry, there are no sharp corners in which the particles
can accumulate;
however, we show that 
confinement effects alone are sufficient to create fluctuation-induced forces.
The magnitude of the attractive force increases with 
increasing run length and can vary over several orders of magnitude. 
For fixed run length,
the force $F$ between the plates as a function of plate spacing $d$ obeys 
$F(d) \propto \exp(-\lambda d)$, 
rather than a power law which is
usually observed in Casimir geometries \cite{1,2,3,6}.  
The attractive force arises
due to the depletion of the
particle density between the plates by a combination of the motion of the
particles along the walls and geometric shadowing, both of which
increase with increasing run length.
We also examine other geometries and show that it is possible 
to control the magnitude of the
forces and 
induce a crossover
from an attractive to a repulsive force. 
For the case of infinite walls,
we analytically derive
the force 
as a function of the distance between the walls.

\begin{figure}
\includegraphics[width=2.0in]{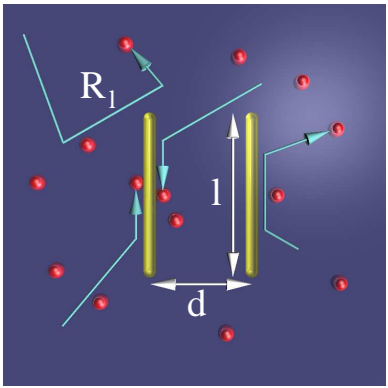}
\caption{
A schematic of the system containing run-and-tumble particles (balls)
with some particle trajectories indicated by lines and arrows.
The run length is $R_l$.
The two parallel walls (bars) are of length $l$ and are separated
by a distance $d$. 
When a particle moves along a wall it imparts a force 
against the wall. 
We measure the net 
force 
$\delta F = |F_{\rm out}|  - |F_{\rm in}|$ 
for varied $R_l$
and $d$, and consider both noninteracting 
and dilute interacting particles.
}
\label{fig:1}
\end{figure}

{\it Simulation-- }   
In Fig.~1 we show a schematic of our system which resides in a 
two-dimensional (2D) simulation box of size $L_x \times L_y$ 
with periodic boundary conditions in the $x$- and $y$-directions.  
Unless otherwise noted,
we take $L_x=120$ and $L_y=60$.
Within the box are $N$ active particles and
two parallel walls of length $l$ 
separated by a distance $d$.  
The run-and-tumble particles
move in a fixed randomly chosen direction during a running time $\tau$ 
before undergoing a tumbling event and running again in a new randomly
chosen direction.
Since we consider the dilute limit where particle-particle 
interactions are rare and clustering does not occur,
for efficiency
we use
event driven (ED) dynamics
simulations. 
The particles run with speed $v=1$, and we neglect particle-particle
interactions.  Instantaneous tumble events occur every $\tau$ time units,
so during a free run the particle moves a run length of
$R_l=\tau v$. 
When a particle encounters a wall, the wall absorbs the 
component of motion perpendicular to the wall:
that is, a particle contacting a wall at an angle $\theta$
moves with velocity $\sin\theta$ along 
the wall and exerts a force $\cos\theta$ on the wall. 
If the particle reaches the end of the wall before tumbling, it returns
to its original swimming direction.
The assumption that run-and-tumble particles move along or
accumulate at walls
has
been experimentally confirmed in ratchet geometry \cite{14} and 
asymmetric gear experiments \cite{32,33} and observed
in simulations \cite{26,27,28}.  
We measure the 
time-averaged 
force $F_{\rm out}$ imparted on the outside edges of the walls 
and $F_{\rm in}$ imparted on the inside edges.  
The net force acting to bring the walls together
or apart is $\Delta F = |F_{\rm out}|-|F_{\rm in}|$; 
we report the net force per particle, $\Delta F/N$.

We also confirmed our results using molecular dynamics (MD) simulations with 
and without particle-particle interactions and find that 
all of our results are robust for the particle densities we consider. 
In the MD simulations the dynamics of particle $i$
are obtained by integrating
the 
overdamped equation of motion
$\eta 
d {\bf R}_{i}/dt = 
{\bf F}^{m}_{i} + {\bf F}^{s}_{i} +  {\bf F}^{b}_{i}. $
Here $\eta = 1.0$ is the damping constant and 
${\bf F}^{m}_{i}$ is the motor force
which has a fixed magnitude $F^m=1$.  
During each run time $\tau$ 
the particle moves a distance
$R_{l} = F^m\tau$ in the absence of interactions. 
The particle-particle steric interactions 
are a stiff-spring repulsion 
given by
${\bf F}^{s}_{i} = \sum^{N}_{j \neq i}k_s(2r_d - |{\bf r}_{ij}|)
\Theta(2r_d -|{\bf r}_{ij}|){\hat {\bf r}}_{ij}$,
where ${\bf r}_{ij} = {\bf R}_{i} - {\bf R}_{j}$, 
${\hat {\bf r}}_{ij} = {\bf r}_{ij}/|{\bf r}_{ij}|$, 
${\bf R}_{i(j)}$ is the location of particle $i(j)$,
$k_s=300$ is the spring constant,
and $r_d=0.5$ is the particle radius.
The particle-barrier interactions 
${\bf F}^{b}_{i}$ 
are of a similar form, 
and are the same as those used in previous
simulations \cite{26,drocco}.
When a particle contacts a wall 
it moves along the wall with the component of its motor force parallel to
the wall and imparts the perpendicular force component
to the wall.

\begin{figure}
\includegraphics[width=3.5in]{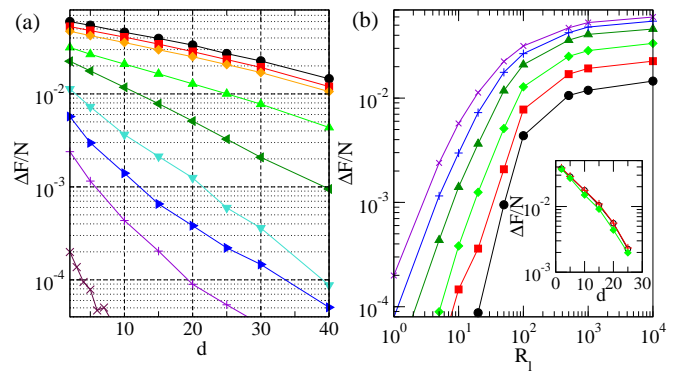}
\caption{
(a) $\Delta F/N$, the attractive force between the two walls, vs 
$d$, the distance between the walls, for ED simulations with
$l = 20$. 
From top to bottom, 
$R_{l} = 10000$, 100, 500, 100, 50, 20, 10, 5, and  1.
The curves can be fit by $\Delta F/N \propto A\exp(-\lambda d)$.
(b) $\Delta F/N$ vs $R_l$ for the same system at
$d = 2$, 5, 10, 20, 30, and $40$ from top to bottom. 
For large $R_{l}$, $\Delta F/N$ saturates.
Inset: $\Delta F/N$ vs $d$ for $l=20$, $R_l=20$, and $N=400$ in 
ED simulations 
($\bullet$) and MD simulations without (+) and with ($\blacklozenge$)
steric particle-particle interactions.
There is only a small shift in $\Delta F/N$ at large $d$ for the 
interacting system.
}
\label{fig:2}
\end{figure}

{\it Results-- }  
In Fig.~2(a) we plot the normalized force $\Delta F/N$ versus 
$d$ for 
ED simulations 
of a 
system with 
$l = 20$ at
run lengths $R_{l} = 10000$, 1000, 500, 100, 50, 20, 5, and  1.   
The magnitude of $\Delta F/N$ increases with increasing $R_{l}$ and we find
$\Delta F(d) \propto A\exp(-\lambda d)$, with $\lambda$ decreasing 
for increasing $R_{l}$. 
At small runs lengths such as $R_l=1$, $\Delta F/N$
becomes very small, and in the limit of
infinitesimal $R_l$ or Brownian particles there is 
no attractive force between the walls. 
In Fig.~2(b) we plot          
$\Delta F$ versus $R_{l}$ for the same system at
$d = 2$, 5, 10, 20, 30, and $40$, 
showing that the attractive force increases with increasing $R_{l}$ 
and saturates at large $R_{l}$. 
To verify the robustness of the results, 
we perform three different types of simulations 
highlighted in the inset of
Fig.~2(b) where we plot
$\Delta F/N$  vs $d$ for systems with $N=400$, $l=20$, and $R_l=20$.
The results of the ED and MD simulations with noninteracting particles are
almost identical.  When steric particle-particle interactions are included
in the MD simulations,
$\Delta F/N$ is nearly unchanged for small $d$ and slightly reduced
at larger $d$, and 
still depends exponentially on $d$. 
This indicates that our results 
are robust 
in the dilute limit.
At higher particle densities, phase separation or clustering becomes
important if steric particle-particle interactions are included
\cite{20,21,22}, a case that we do not consider in this work.  

\begin{figure}
\includegraphics[width=3.5in]{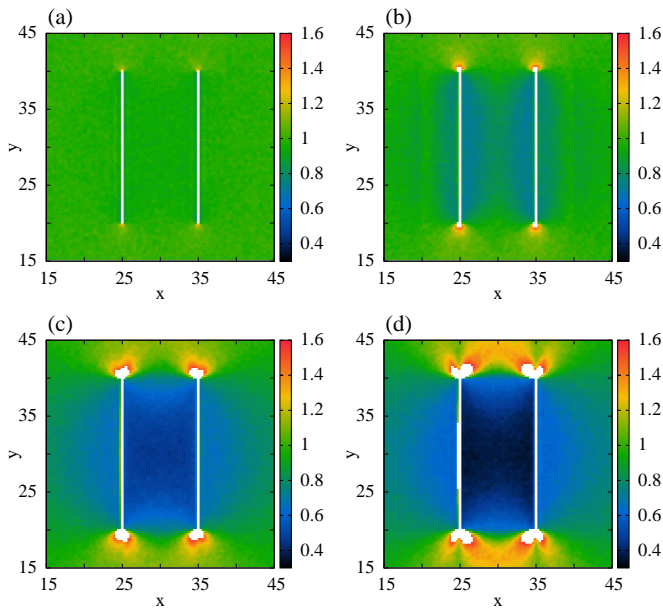}
\caption{
$\rho({\bf r})$, the
spatial distribution of the average particle density, for the system in
Fig.~2(a) with 
$L_x = 60$, $l = 20$, and $d=10$ at
(a) $R_{l} = 1$, (b) $R_{l} = 5$, (c) $R_{l} = 20$, and (d) $R_{l} = 50$. 
$\rho$ becomes increasingly depleted between the plates
with increasing $R_l$,
leading to a larger net attractive force between the walls. 
}
\label{fig:3}
\end{figure}

In order to better understand the origin of the attractive force,
in Fig.~3 we plot 
$\rho({\bf r})$, the spatial distribution of the average particle density,
from ED simulations with 
$L_x = 60$, $l = 20$, and $d =10$ for different values of $R_l$.
We use the 
normalization that $\rho=1.0$
in 
a homogeneous system without walls. 
For short run lengths $R_{l} = 1$ 
in Fig.~3(a), $\rho \approx 1$
throughout the sample
and there is little to no attractive force
between the walls.
Although $\rho>1$
immediately adjacent to the walls, 
the density of particles on the inside and
outside edges of the walls is virtually identical.
At $R_{l} = 5$, shown in Fig.~3(b), two trends
start to appear. There is a depletion region of 
lower $\rho$
near the walls which is most
pronounced at their centers.
In addition, the higher 
$\rho$ along the walls produces
high density spots at the wall ends where
the particles escape from the walls and disperse back into the bulk.
The depletion of 
$\rho$ between the walls is 
similar to a ray optics shadow effect caused by the finite run length $R_l$. 
Particles exerting forces on the outer edges of the walls can originate from
a broad region.  In contrast, particles exerting forces on the inner walls 
originate from a much smaller region.  The resulting shadowing effect
depletes $\rho$ between the walls.
As $d$
decreases, the difference in size between the inner and outer regions grows,
leading to an increased shadow effect and 
hence a stronger $\Delta F$ for smaller $d$.    
At $R_{l} = 20$ in Fig.~3(c), the shadowing effect becomes more 
pronounced and 
$\rho$ is strongly reduced
between the walls,
while in Fig.~3(d), for $R_{l} =50$
$\rho$ between the walls is even further reduced and   
the density distribution at the ends of the walls develops additional 
features due to interference effects. 
These results show that Casimir attractive forces 
can arise in active matter systems and that they can be controlled 
by modifying the run length
or persistence length of the swimming particles. 

\begin{figure}
\includegraphics[width=3.5in]{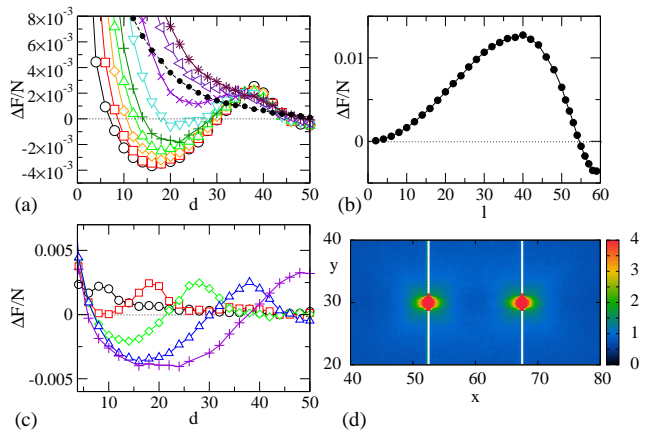}
\caption{
(a) $\Delta F/N$ vs $d$
for a system with 
$R_{l} = 40$ 
at $l = 20$ ($\bullet$), 40 (*), 45 ($\triangleleft$), 50 (x), 
53 ($\bigtriangledown$),
55 (+), 56 ($\bigtriangleup$), 57 ($\Diamond$), 
58 ($\Box$)
and $59$ ($\bigcirc$). 
For  $l \geq 55$ there is a regime where the force becomes negative. 
(b) $\Delta F/N$ vs $l$ for fixed $d = 15$ and $R_{l}  = 40$. 
Initially, $\Delta F/N$ is attractive and
increases before reaching
a maximum near $l = 40$ and then becoming negative at larger $l$.      
(c) $\Delta F/N$ vs $d$ for a system with $l = 59$
at $R_{l} = 10$ ($\bigcirc$), 20 ($\Box$), 30 ($\Diamond$), 
40 ($\bigtriangleup$) and $50$ (+). 
(d) 
$\rho({\bf r})$
for a system with 
$l = 58$ at $d = 15$.
}
\label{fig:4}
\end{figure}

{\it Reverse Casimir Effect-- }
We next consider the effect of changing $l$, the wall length.
Since we have periodic boundary conditions, 
at large $l$ the system is better described as consisting of apertures
of width $w=L_y-l$,
as shown in Fig.~4(d) for a system with 
$l = 58$. 
In Fig.~4(a) we plot $\Delta F/N$ versus $d$ for a system with
$R_{l} = 40$ 
at $l = 20$, 40, 45, 50, 53, 55, 56, 57, 58, and $59$. 
For 
$l<55$, the force is attractive for all $d$.
For $l \geq 55$, there is a range of $d$
for which $\Delta F/N$  becomes negative, 
while at larger $d$
the force becomes positive again. 
The magnitude of the negative force 
depends on $l$, with the largest negative forces appearing for the smallest
values of $w$.
In Fig.~4(b) we plot $\Delta F/N$ versus $l$ 
for a system with $d = 15$ and $R_{l} = 40$. 
Here $\Delta F/N$ is nearly zero 
at small $l$ and increases with increasing $l$ up to
$l \approx 40$ due to 
the 
shadowing effect. 
For $l > 40$, $\Delta F/N$ decreases rapidly and becomes negative,
competing with the shadowing effects that tend to increase the attractive 
force.  The drop arises from
a particle trapping effect that occurs as the system 
enters the aperture limit.
The particles moving along an outside wall return to the
bulk upon reaching the wall end when $l$ is small, but for large $l$
they instead are trapped by the interior region,
raising the interior density and producing a net repulsive force between
the two walls.
The trapping effect becomes more prominent with decreasing $w$.
In Fig.~4(a), where $R_l=40$, for $l \geq 53$ and 
$d < 30$ the particles in the interior region spend most of their time
running along the 
interior walls, 
producing a repulsive force.
For $d > 30$ the particles have a chance to turn away from the inside wall
during their run and avoid striking it,
so the force becomes attractive again. 

In Fig.~4(c) we plot $\Delta F/N$ versus $d$ 
for systems with different 
$R_l$.
For the short run length of $R_{l}= 10$, $\Delta F/N$
is always positive, but as $R_{l}$ increases, the force drops below zero
over a region that increases with increasing $R_l$, 
and returns to positive values at higher values of $d$.
The resulting shape of $\Delta F/N(d)$ resembles an atomic interaction
force curve, with a 
stable characteristic distance determined by the point at which
$\Delta F/N$ crosses zero
with positive slope, suggesting that freely moving walls could
be stabilized at this spacing.
The spacing can be
adjusted by varying $R_l$ in order to move the zero point of $\Delta F/N$ to
the desired distance, opening the possibility that active matter could be
used to self-assemble passive objects not merely to the aggregation point of
zero spacing, but also to a finite spacing set by $R_l$.
These results show that by varying the geometry it is possible
to achieve detailed control over the magnitude of the fluctuation-induced
forces.      

\begin{figure}
\includegraphics[width=3.5in]{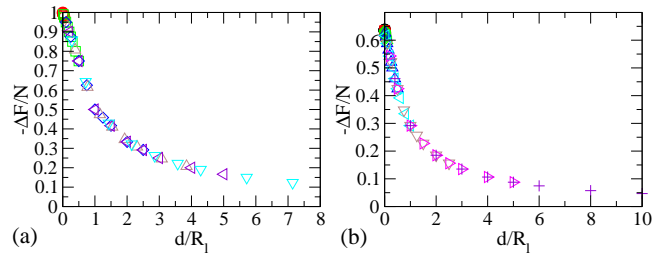}
\caption{
The outward force $-\Delta F/N$
vs $d/R_{l}$ for a system with infinite walls 
where the active particles are confined to move only between
the walls from 
ED simulations.
The particle density between the walls is held constant.  
(a) 1D systems with $R_l$=1000 ($\bigcirc$), 100 ($\Box$),
20 ($\Diamond$), 13 ($\bigtriangleup$), 10 ($\triangleleft$),
and 7 ($\bigtriangledown$). The curves can be collapsed
to the functional form $\Delta F(d) = R_{l} (R_{l} + d)^{-1}$.
(b) 2D systems with $R_l$=10000 ($\bigcirc$), 1000 ($\Box$),
500 ($\Diamond$), 100 ($\bigtriangleup$), 50 ($\triangleleft$),
20 ($\bigtriangledown$), 10 ($\triangleright$), and 5 (+).
A similar data collapse occurs.
}
\label{fig:5}
\end{figure}

{\it Infinite Walls-- }
We have also considered the case of infinite walls that confine the particles
to an interior region, producing only an outward force.
Such a system could be created by trapping bacteria or active matter particles
in a confined geometry and measuring the force exerted on the walls. 
We consider both one-dimensional (1D) and 2D infinite wall systems 
and measure the force for different wall spacings $d$, with the 
interior particle
density held fixed.
In the limit of Brownian particles,
$\Delta F$ would be independent of $d$.
In Fig.~5(a) we plot the outward force 
$-\Delta F/N$ versus 
$d/R_l$ 
for a 1D system. 
When $l$ is infinite, the only length scales are 
$d$ and 
$R_l$.
The fraction of time a particle spends traveling along the 
walls is unchanged if both these lengths are scaled by the same factor, 
and so the force must be a function of the ratio $d/R_l$. 
Fig.~5 shows that
the curves for various $R_l$ can be collapsed to the form 
$F(x) = 1/(1 + x)$ with $x \equiv d/R_l$, 
which can be derived from elementary considerations. 
For simplicity, consider integer $x$
and measure length in units of $R_l$ and time in units of $\tau$.
The particle moves in 1D along $y$ with walls located
at $y=0$ and $y=x$. 
During each run time, the particle moves a distance $\pm 1$ along $y$.
To calculate the relative amount of time 
a particle spends at a wall, we note that 
a particle emerges into the bulk 
by tumbling
away from a wall; taking this to be the $y=0$ wall, the particle 
takes one step to $y=1$. We then need to calculate the expected 
time $t_b$ required
for the particle to either return to $y=0$ or 
reach $y=x$.  This is the ``gambler's ruin'' problem, 
with the well-known solution of $(x-1)$. Counting the first step into 
the bulk, the expected number of steps in the bulk between 
wall motions is just $x$, so $t_b=x$.
To determine $t_w$, the amount of time 
the particle spends at the wall after reaching it,
we note that 
each tumble provides a 50:50 chance of escaping the wall, 
giving a probability $p=\frac{1}{2}$ that $t_w=0$,
$p=\frac{1}{4}$ that $t_w=1$, and so on. The expected value of $t_w$
is easily found to be $t_w=1$. Therefore, the ratio 
of $t_w$ to the total time is $t_w/(t_w+t_b)=1/(1 + x)$. 
This equals the time-averaged force on the walls 
since in the 1D scenario, the force on a 
wall is 1 when a particle is present at the wall and 0 otherwise.
For noninteger $x$, it is not too difficult to derive a more 
general formula $\frac{2-\frac{x}{{\rm ceil}(x)}}{1+{\rm ceil}(x)}$ where 
${\rm ceil}(x)$ is the smallest integer $\geq x$.

In Fig.~5(b) we plot $-\Delta F/N$ for a 2D system.
Here we could not derive a fully analytic solution; however, we
obtain a fit of $F(x)= a/(1 + bx)$ with $a = 2/\pi$ and $b \approx 1.25$.
The value for $a$ is exact
and arises
because in 2D 
all angles of incidence are possible 
rather than just normal incidence. Since the angle distribution is uniform, 
the resulting force on the wall is lessened by a factor of 
$\frac{1}{\frac{\pi}{2}-0}\int_{0}^{\pi /2} \cos\theta \ {\rm d}\theta 
= \frac{2}{\pi}$. We cannot calculate $b$ exactly, but it can 
be understood heuristically: compared
to the above 1D derivation, 
the particle spends more time in the bulk 
because only the $y$ component of its motion 
moves it towards a wall. Therefore the effective step size 
is reduced from 1 and $t_b$ is increased, giving $b>1$.
When steric particle-particle interactions are included, 
the forces on the walls are reduced; however, in the dilute limit these
results should hold and should be readily testable in experiments.              

{\it Summary-- } 
We have shown that in a Casimir geometry of 
two parallel walls placed in a bath of run-and-tumble
active particles, a robust attractive force arises between the walls due
to a combination of the particles moving along the walls 
and a geometric shadowing effect that 
depletes the particle density between the walls.  
The depletion becomes more pronounced for increasing particle run length or
reduced wall spacing, increasing the magnitude of the force.
Our results are robust against the inclusion of steric particle-particle
interactions in the dilute limit.
We also find that 
for other geometries such as two walls containing
small apertures,
a particle trapping effect can produce repulsive forces between the walls.
Our results show that active matter systems can 
exhibit a rich variety of fluctuation-induced forces
between objects, and these effects 
may be useful for applications such 
as self assembly or particle transport.

\acknowledgments
We thank D. Dalvit and L. Lopatina for useful discussions. 
This work was carried out under the auspices of the 
NNSA of the 
U.S. DoE
at 
LANL
under Contract No.
DE-AC52-06NA25396.

\end{document}